\documentclass[aps,apl,twocolumn]{revtex4}

\usepackage{amsmath}
\usepackage{amssymb}
\usepackage[dvipdfmx]{graphicx}  
\usepackage[dvipdfmx]{color}
\usepackage{verbatim}   
\usepackage{color}      

\newcommand{\pro}[2]{\langle{#1}|{#2}\rangle}
\newcommand{\bra}[1]{\langle{#1}|}
\newcommand{\ket}[1]{|{#1}\rangle}

\begin{document}

\title{Control limit on quantum state preparation under decoherence}
\author{Kohei Kobayashi}
\author{Naoki Yamamoto}
\affiliation{Department of Applied Physics and Physico-Informatics, Keio University, 
Hiyoshi 3-14-1, Kohoku, Yokohama, 223-8522, Japan}
\date{\today}

\begin{abstract}
Quantum information technologies require careful control for generating and preserving 
a desired target quantum state. 
The biggest practical obstacle is, of course, decoherence. 
Therefore, the reachability analysis, which in our scenario aims to estimate the distance 
between the controlled state under decoherence and the target state, is of great importance 
to evaluate the realistic performance of those technologies. 
This paper presents a lower bound of the fidelity-based distance for a general open 
Markovian quantum system driven by the decoherence process and several types of 
control including feedback. 
The lower bound is straightforward to calculate and can be used as a guide for choosing 
the target state, as demonstrated in some examples. 
Moreover, the lower bound is applied to derive a theoretical limit in some quantum metrology 
problems based on a large-size atomic ensemble under control and decoherence. 
\end{abstract}

\maketitle


\section {Introduction}

There is no doubt that a carefully designed control plays a key role in quantum information 
science. 
The open-loop (i.e., non-feedback) control theory 
\cite{Rabitz 1993,Viola 1999,Khaneja 2003,D'Alessandro Book, Jacobs 2007,Jacobs 2009,
Montangero PRL 2011, Laflamme 2016} 
offers several powerful means, for example, for implementing an efficient quantum gate operation. 
The measurement-based feedback (MBF) 
\cite{Stockton 2004,Handel 2005,Geremia 2006,Yanagisawa 2006,Yamamoto 2007,
Molmer 2007,Mirrahimi 2007,Bouten 2009} and reservoir engineering including 
coherent feedback 
\cite{Poyatos 1996,Schirmer 2010,Yamamoto 2014,Vitali2014,Clerk2016,
Combes 2017,Ferraro 2018} 
are also well-established methodologies that can be used for generating and protecting 
a desired quantum state. 
Remarkably, many notable experiments realizing those control techniques have been 
demonstrated 
\cite{Haroche 2011,Siddiqi 2012,Lehnert 2013,Siddiqi 2016,Thompson 2016,Devoret 2016}, 
which at the same time show that the actual control performance in the presence of 
decoherence is sometimes far away from the ideal one. 
Therefore the {\it reachability analysis} is essential to evaluate the practical effectiveness of 
those control methods; 
that is, it is important to quantify how close the controlled quantum state can be steered to 
or preserved at around a target state under decoherence. 
Note that the reachability characteristic determines a lower bound of the time required for 
performing a desired state transformation via control \cite{Lloyd 2014}.

The reachability analysis found in the literature is usually based on simulations, 
which numerically investigate how much the ideal state control is disturbed by 
decoherence; 
for example, generation of a nano-resonator superposition state via open-loop control 
\cite{Jacobs 2007,Jacobs 2009}, an optical Fock state via MBF 
\cite{Geremia 2006,Molmer 2007}, and an opto-mechanical cat state via reservoir 
engineering \cite{Vitali2014,Ferraro 2018}. 
The optimal control method is also often used, which numerically designs a time-dependent 
control input for steering the state closest to the target under decoherence 
\cite{Khaneja 2003,Koch 2016}. 
However, these computational approaches do not give us deep insight into basic 
questions for quantum engineering, e.g., what state should be targeted, what the limit of 
realistic state preparation is, and what the desired structure of open quantum systems 
under given decoherence is. 
A few exceptions are found for specific types of open-loop control \cite{Khaneja pnas 2003} 
and MBF \cite{Guo 2010,Guo 2013}, but there has been no unified approach. 
Also the controllability property, which is a stronger notion than the reachability, can be 
analytically investigated using the Lie algebra 
\cite{Altafini 2003,Khaneja 2009,Kurniawan 2009,Kurniawan 2012,Dirr 2012,Yuan 2013}; 
but it does not quantify the distance to the target and thus does not answer the above 
questions.

The main contribution of this paper is to present a limit for reachability, applicable for 
a general open Markovian quantum system driven by the decoherence process and several 
types of control including the open-loop and MBF controls and reservoir engineering; 
more precisely, we give a lower bound of the fidelity-based distance between a given 
target state and the controlled state under decoherence. 
This lower bound is straightforward to calculate, without solving any equation. 
Also thanks to its generic form, the lower bound gives a characterization of target states 
that are largely affected by the decoherence, and thereby provides us a useful guide 
for choosing the target, as demonstrated in some examples. 
Moreover, as ``a deep insight into quantum engineering", the lower bound is used to 
derive a theoretical limit in quantum metrology; 
for a typical large-size atomic ensemble under control and decoherence, the fidelity to 
the target (the Greenberger-Horne-Zeilinger (GHZ) state or a highly entangled Dicke state) 
must be less than 0.875, regardless of the control strategy.


\section{The control limit}

\subsection{Controlled quantum dynamics}

We begin with a simplified setting of open-loop control and reservoir engineering; 
the quantum state $\bar{\rho}_t$ obeys the Markovian master equation 
\begin{equation}
\label{linear ME}
      \frac{d\bar{\rho}_t}{dt}
         = -i [u_t H, \bar{\rho}_t] + \mathcal{D}[L] \bar{\rho}_t + \mathcal{D}[M] \bar{\rho}_t, 
\end{equation}
where $H$ is a system Hamiltonian. 
$L$ and $M$ are Lindblad operators, and hence 
$\mathcal{D} [A]\bar{\rho} = A\bar{\rho} A^\dagger - A^\dagger A\bar{\rho}/2 
- \bar{\rho} A^\dagger A/2 $. 
Here $L$ represents the uncontrollable ``L"indblad operator corresponding to the 
decoherence, while $M$ can be engineered; in particular in the MBF setting $M$ 
represents the probe for ``M"easurement. 
The standard open-loop control problem is to design a time-dependent sequence 
$u_t$ that steers $\bar{\rho}_t$ toward a target state, under $M=0$. 
Also the standard reservoir engineering approach aims to design $M$, with constant 
$u_t$, so that $\bar{\rho}_t$ autonomously converges to a target.

The MBF control setting can also be included in the theory. 
In this case the quantum state $\rho_t$ conditioned on the measurement record $y_t$ 
obeys the following stochastic master equation (SME) 
\cite{Wiseman Book,Jacobs Book,NY Book}: 
\begin{equation}
\label{SME}
          d\rho_{t} = -i[u_{t}H, \rho_{t}]dt+\mathcal{D}[L]\rho_{t}dt
                           +\mathcal{D}[M]\rho_{t}dt+\mathcal{H}[M]\rho_{t}dW_{t}, 
\end{equation}
where $dW_{t}=dy_{t}-{\rm Tr}[(M+M^{\dagger})\rho_{t}]dt$ is the Wiener increment 
representing the innovation process based on $y_{t}$, and 
$\mathcal{H}[A]\rho = A\rho+\rho A^\dagger - {\rm Tr}[(A+A^\dagger)\rho]\rho$. 
The goal of MBF is to design the control signal $u_t$ as a function of $\rho_t$, to 
achieve a certain goal. 
In particular, if $L=0$ and $M=M^\dagger$, there are several types of MBF control that 
selectively steer the state to an arbitrary eigenstate of $M$. 
In this paper, we focus on the unconditional state $\bar{\rho}_t={\mathbb E}(\rho_t)$, 
which is the ensemble average of $\rho_t$ over all the measurement results. 
Then due to ${\mathbb E}(W_t)=0$, Eq.~\eqref{SME} leads to the following master equation:
\begin{equation}
\label{nonlinear ME}
      \frac{d\bar{\rho}_t}{dt}
         = -i [H, {\mathbb E}(u_t\rho_t)] + \mathcal{D}[L] \bar{\rho}_t 
                  + \mathcal{D}[M] \bar{\rho}_t.
\end{equation}
Note that now $u_t$ is a function of $\rho_t$, and thus Eq.~\eqref{nonlinear ME} is not 
a linear equation with respect to $\bar{\rho}_t$. 
In the open-loop control or reservoir engineering setting, meaning that $u_t$ is 
independent of $\rho_{t}$,  then Eq.~\eqref{nonlinear ME} is reduced to the linear 
equation \eqref{linear ME} due to ${\mathbb E}(u_t\rho_t)=u_t\bar{\rho}_t$.


\subsection{Main result}

The control goal is to minimize the following cost function: 
\begin{equation}
\label{cost}
     J_t = 1 - \bra{\psi} \bar{\rho}_t \ket{\psi},
\end{equation}
where $\ket{\psi}$ is the target pure state and $\bar{\rho}_t$ is the controlled state 
obeying Eq.~\eqref{linear ME} or \eqref{nonlinear ME}; 
hence, $J_t$ represents the fidelity-based distance of $\bar{\rho}_t$ from the target. 
Under the presence of decoherence term ${\mathcal D}[L]$, in general it is impossible to 
deterministically achieve $J_t=0$ at some time $t$. 
The main result of this paper is to provide a lower bound of the cost in an explicit 
form as follows. 
The proof is given in Appendix~A. 
\\

{\it Theorem~1}: 
The cost \eqref{cost} has the following lower bound at the steady state: 
\begin{equation}
       J_\infty \geq J_{\ast}(\ket{\psi})
           = \left(\frac{{\mathcal E}}{{\mathcal A}+{\mathcal U}} \right)^2,
\end{equation}
where ($\|\ket{\psi}\|^2=\pro{\psi}{\psi}$ is the Euclidean norm) 
\begin{align*}
      {\mathcal A} &= \sqrt{2}\bigl( 
                  \| L^\dagger \ket{\psi} \|^2 + \| L^\dagger L \ket{\psi} \| 
\\
         & \hspace{1cm} + \| M^\dagger \ket{\psi} \|^2 + \| M^\dagger M\ket{\psi} \|   \bigr), 
\\
     {\mathcal U} & = 2 \bar{u} \sqrt{ \bra{\psi}H^2\ket{\psi} - \bra{\psi}H\ket{\psi}^2 }, 
     ~~\bar{u}=\max\{ |u_t| \}, 
\\
     {\mathcal E} &=  \| L \ket{\psi} \|^2 - | \bra{\psi} L \ket{\psi} |^2 
                               + \| M \ket{\psi} \|^2 - | \bra{\psi} M \ket{\psi} |^2. 
\end{align*}
Moreover, if $J_{t_0}\geq J_*$ for an initial state $\bar{\rho}_{t_0}$, then 
$J_t \geq J_*$ holds for all $t\in[t_0, \infty)$. 
\\

That is, $J_*$ gives a limit on how close the controlled quantum state can be steered to 
or preserved at around a target state under decoherence. 
Below we list some notable general features of $J_*$. 

(i) The theorem is applicable to a general Markovian open quantum systems 
driven by several types of control including the MBF and reservoir engineering. 

(ii) The result can be extended to the case where the system is subjected to 
multiple environment channels, measurement probes, and control Hamiltonians, as 
long as the dynamical equation can be validly described as an extension of 
Eq.~\eqref{linear ME} or \eqref{SME}; see Appendix~B. 

(iii) $J_*$ is directly computable, once the system operators and the target state $\ket{\psi}$ 
are specified; it is not necessary to solve any equation. 

(iv) $J_*$ is a monotonically decreasing function of the control magnitude $\bar{u}$. 

(v) If $\ket{\psi}$ moves away from the eigenstates of $L$ and $M$, then $J_*$ becomes 
bigger. 
Conversely, $J_*=0$ if and only if $\ket{\psi}$ is identical to a common eigenvector of $L$ 
and $M$.

Importantly, $J_*$ can be used to characterize a target state that is possibly easy to approach 
by some control, under a given decoherence. 
That is, a state $\ket{\psi}$ with relatively small value of $J_*$ might be a good candidate as 
the target, although in general $J_*$ is not achievable. 
Conversely, we can safely say that the state $\ket{\psi}$ with a relatively large value of $J_*$ 
should not be assigned as the target. 
In what follows we study some typical control problems, with special attention to this point.


\section{Examples}

\subsection{Qubit}

The first example is a qubit such as a two-level atom, consisting of $\ket{e}=[1, 0]^{\top}$ 
and $\ket{g}=[0, 1]^{\top}$. 
Let the target be a pure qubit state 
\begin{equation}
\label{qubit target}
     \ket{\psi}=[\cos{\theta}, ~e^{i\varphi}\sin{\theta}]^{\top}, 
     ~(0\leq \theta <\pi/2, \ 0\leq \varphi<2\pi). 
\end{equation}
Here we consider the following operators:
\[
     H = \sigma_{y}, ~M=\sqrt{\kappa}\sigma_{z}, ~L=\sqrt{\gamma}\sigma_{-}, 
\]
where $\sigma_y=i(\ket{g}\bra{e}-\ket{e}\bra{g})$, $\sigma_z=\ket{e}\bra{e}-\ket{g}\bra{g}$, 
and $\sigma_{-}=\ket{g}\bra{e}$. 
This is a typical MBF control setup \cite{Handel 2005,Siddiqi 2012,Lehnert 2013,Siddiqi 2016}; 
$M$ represents the engineered dispersive coupling between the qubit and the probe, which 
enables us to 
continuously monitor the qubit state by measuring the probe output and thereby perform a 
MBF control via the Hamiltonian $u_t H$. 
Ideally (i.e., if $\gamma=0$), this MBF realizes deterministic and selective steering of 
the qubit state to $\ket{e}$ or $\ket{g}$. 
However in practice this perfect control is not allowed due to decoherence 
$L=\sqrt{\gamma}\sigma_{-}$, i.e., energy decay from $\ket{e}$ to $\ket{g}$ 
for a two-level atom. 
In this setup, the lower bound $J_*$ is calculated as 
\[
       J_{\ast}=\frac{1}{2}\left[
                        \frac{\kappa \sin^2 2\theta + \gamma\cos^4\theta}
                               {2\kappa+\gamma(\sin^2\theta+\cos\theta)+
                                   \bar{u} \sqrt{2-2\sin^2 2\theta \sin^2\varphi }} \right]^2.
\]
A detailed derivation of this expression (and that of $J_*$ in the other examples) 
is provided in Appendix~D.

\begin{figure}[htbp]
\includegraphics[width=8.8cm]{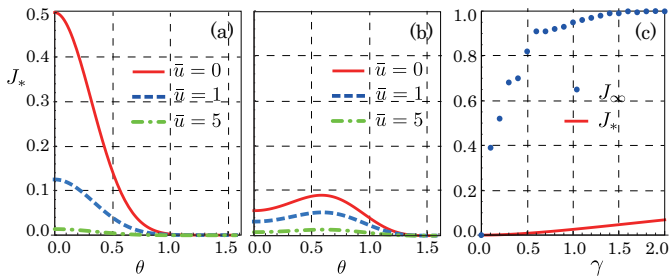}
\caption{The lower bound $J_*$ as a function of $\theta$ for (a) $\kappa=0$ and 
(b) $\kappa=1$, in units of $\gamma=1$. 
(c) Plot of $J_*$ and $J_\infty$ with $u_t$ a special type of MBF control input, for the case 
$\ket{\psi}=\ket{e}$ and $(\kappa, \bar{u})=(1,1)$. 
}
\end{figure}

First, we set $\kappa=0$; 
this is the case where the system obeys the master equation 
$d\bar{\rho}_t/dt=-i[u_t\sigma_y, \bar{\rho}_t] 
+ {\mathcal D}[\sqrt{\gamma}\sigma_-]\bar{\rho}_t$ driven by 
the open-loop control input $u_t$ satisfying $|u_t|\leq \bar{u}$ 
\cite{Altafini 2003,Schirmer 2010,Yuan 2013,Laflamme 2016}. 
Figure~1(a) shows the above lower bound $J_{\ast}$ in units of $\gamma=1$, for the target 
satisfying $\varphi=0$. 
Clearly, $J_*$ takes the maximum at $\ket{\psi}=\ket{e}$ and zero at $\ket{\psi}=\ket{g}$ 
for each $\bar{u}$, implying that $\ket{e}$ is the most difficult state to approach, while 
$\ket{g}$ could be stabilized exactly; 
in fact these implications are true, as can be analytically verified by solving the above master 
equation. 
On the other hand if $\kappa=1$, as depicted in Fig.~1(b), $J_{\ast}$ at around $\theta=0$ 
remarkably decreases compared to the case $\kappa=0$. 
This is reasonable because the dispersive interaction represented by $M=\sqrt{\kappa}\sigma_z$ 
enables us to perform a MBF control that deterministically stabilizes $\ket{e}$ if $\gamma=0$. 
As a consequence, $J_*$ takes the maximum at around $\theta=0.6$, meaning that a 
superposition is the most difficult state to reach.

It is also worth comparing $J_*$ to the actual distance $J_\infty$ achieved by a special 
type of MBF. 
We particularly take the method proposed in \cite{Mirrahimi 2007} and compute 
$J_\infty$ by averaging 300 sample points of conditional state $\rho_t$ at steady 
state, for the case $\ket{\psi}=\ket{e}$ and $(\kappa, \bar{u})=(1,1)$ with several value 
of $\gamma$; see Appendix~C for the details. 
Figure~1(c) shows that the gap between $J_{\ast}$ and $J_\infty$ is large and hence 
$J_{\ast}$ is not a tight lower bound in this case; 
but one could take another control strategy to reduce the gap and eventually prepare a 
state close to $\ket{e}$.


\subsection{Two-qubits}

Here we study a two-qubits system under decoherence. 
First let us focus on the following Bell states, which are of particular use in the scenario 
of quantum information science \cite{Nielsen Book}: 
\begin{equation*}
     \ket{\Phi^{\pm}}=\frac{1}{\sqrt{2}}\left(\ket{g,g}\pm\ket{e,e}\right), ~
     \ket{\Psi^{\pm}}=\frac{1}{\sqrt{2}}\left(\ket{g,e}\pm\ket{e,g}\right).
\end{equation*}
The question here is which Bell state is the best one accessible by any open-loop control 
(hence assume $M=0$) \cite{Laflamme 2016}; 
as seen in the qubit case, the lower bound $J_*$ gives us a rough estimate of the answer. 
We particularly consider the collective decay process modeled by 
$L=\sqrt{\gamma}(\sigma_{-}\otimes I+I\otimes\sigma_{-})$. 
Then, for the case $\ket{\Phi^+}$, we have 
$\mathcal{E}=\|L\ket{\Phi^+}\|^2-|\bra{\Phi^+}L\ket{\Phi^+}|^2=\gamma$ and 
$\mathcal{A}=\sqrt{2}(\|L^\dagger\ket{\Phi^+}\|^2+\|L^{\dag}L\ket{\Phi^+}\|)
=(2+\sqrt{2})\gamma$. 
Hence, together with the other Bell states, the lower bounds are calculated as 
\begin{equation*}
     J_*(\ket{\Phi^\pm})=\frac{\gamma^2}{[(2+\sqrt{2})\gamma+\mathcal{U}]^2}, ~
     J_*(\ket{\Psi^+})=\frac{4\gamma^2}{(4\sqrt{2}\gamma+\mathcal{U})^2}, 
\end{equation*}
and $J_*(\ket{\Psi^-})=0$. 
Here we assumed that, for each case of the Bell state, an appropriate control Hamiltonian 
$H$ is chosen so that the same magnitude of control, $\mathcal{U}$, appears in the 
expression of $J_*$ for fair comparison. 
Hence, the Bell states, which ideally have the same amount of entanglement, have different 
reachability properties under realistic decoherence. 
Clearly, in our case $\ket{\Psi^-}$ is the best target state; 
this is identical to the dark state of $L$ and is indeed reachable. 
Also $J_*(\ket{\Phi^\pm})<J_*(\ket{\Psi^+})$ holds for all $\gamma$ and $\mathcal{U}$, 
showing that $\ket{\Psi^+}$ is the most fragile Bell state under the collective decay 
process. 
On the other hand, if each qubit experiences a local decay, which is modeled as 
$L_{1}=\sqrt{\gamma}\sigma_{-}\otimes I$ and $L_{2}=\sqrt{\gamma}I\otimes \sigma_{-}$ 
rather than the global decay $L=\sqrt{\gamma}(\sigma_{-}\otimes+I\otimes \sigma_{-})$, 
then we have $J_{\ast}=\gamma^{2}/\{(2+\sqrt{2})\gamma+\mathcal{U}\}^{2}$ for all Bell 
states. 
That is, in this case there is no difference between the Bell states, in view of the reachability 
property.

It is also interesting to see the case of MBF \cite{Yamamoto 2007,Wallraff 2010}. 
In particular we limit the system to a pair of symmetric qubits, which is identical to a qutrit 
composed of three distinguishable states $\ket{E}=[1, 0, 0]^{\top}$, $\ket{S}=[0, 1, 0]^{\top}$, 
and $\ket{G}=[0, 0, 1]^{\top}$. 
Note that $\ket{S}$ corresponds to the entangled state between two qubits. 
Here we limit the target state to the following real vector:
\[
       \ket{\psi} = [\sin(\theta/2) \cos(\varphi/2), \, 
                          \cos(\theta/2), \, 
                          \sin(\theta/2)\sin(\varphi/2) ]^\top, 
\]
where $0\leq \theta, \varphi\leq \pi$. 
The MBF setup considered here is given by 
\begin{equation*}
     H=J_y, \ M=\sqrt{\kappa}J_z, \ L=\sqrt{\gamma}J_-,
\end{equation*}
where $J_y=i(\ket{S}\bra{E}+\ket{G}\bra{S})/\sqrt{2}+{\rm h.c.}$, 
$J_z=\ket{E}\bra{E}-\ket{G}\bra{G}$, and $J_-=\sqrt{2}(\ket{S}\bra{E}+\ket{G}\bra{S})$. 
The continuous measurement through the system-probe coupling represented by $M$, 
ideally, induces the probabilistic state reduction to $\ket{E}$, $\ket{S}$, or $\ket{G}$. 
The decoherence process $L$ represents the ladder-type decay 
$\ket{E}\rightarrow \ket{S} \rightarrow \ket{G}$. 
In this setting the lower bound $J_*(\ket{\psi})$ can be explicitly calculated as a function 
of $(\theta, \varphi)$ and is illustrated in Fig.~2. 
As in the qubit case, $J_*(\ket{\psi})$ takes the maximum at $\ket{E}$ when 
$\kappa=0$ (Fig.~2(a)), but $J_*(\ket{E})$ can be drastically decreased by taking a 
non-zero $\kappa$ (Fig.~2(b)); 
that is, the measurement enables us to combat with the decoherence and have chance to 
closely approach to $\ket{E}$ via a MBF. 
However, this strategy does not work for the case of $\ket{S}$, because $J_*(\ket{\psi})$ is 
independent of $\kappa$ at $\theta=0$. 
In general, if the target $\ket{\psi}$ is an eigenstate of $M=M^\dagger$ with small eigenvalue, 
then the term related to $M$ takes a small value as well in ${\mathcal A}$ and zero in 
${\mathcal E}$. 
In particular for the dark state satisfying $M\ket{\psi}=0$, $J_*(\ket{\psi})$ is independent 
on $M$, hence in this case the measurement does not at all help to decrease $J_*$. 
Conversely, for an eigenstate of $M$ with a large eigenvalue, i.e., a bright state $\ket{E}$ 
in our case, the term related to $M$ in ${\mathcal A}$ takes a large number and eventually 
$J_*$ becomes small, implying that we could closely approach to such a state via some 
MBF control even under decoherence.

\begin{figure}[htbp]
\includegraphics[width=8.6cm]{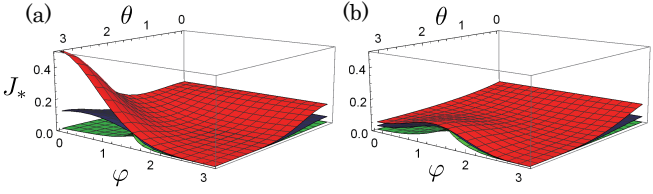}
\caption{
The lower bound $J_*(\ket{\psi})$ as a function of $(\theta, \varphi)$ for (a) $\kappa=0$ and 
(b) $\kappa=1$, in the units of $\gamma=1$. 
In both cases the curved surface corresponds to $\bar{u}=0, 1, 5$ from top to bottom.}
\end{figure}


\subsection{Atomic ensemble}

Next we study an ensemble composed of $N$ identical atoms. 
The basic operators for describing this system are the angular momentum operator 
$J_{i}$ ($i=x,y,z$) satisfying e.g., $[J_x, J_y]=iJ_z$, the magnitude ${\bf J}^2=J_x^2+J_y^2+J_z^2$, 
and the ladder operator $J_-=J_x-iJ_y$. 
Here we focus on the Dicke states $\ket{l, m}$, which are the common eigenstates of 
$J_z$ and ${\bf J}^2$ defined by $J_z\ket{l, m}=m\ket{l, m}$ and 
${\bf J}^{2}\ket{l,m}=l(l+1)\ket{l,m}$ where $|m|\leq l \leq N/2$ \cite{Dicke state Ref Book}. 
Recall, for $N$ even, that $\ket{N/2, N/2}$ corresponds to the coherent spin state (CSS) 
$\ket{\uparrow}^{\otimes N}$, i.e., the separable state with all the spins pointing along the 
$z$ axis, while $\ket{N/2, 0}$ is highly entangled.

It was proven in \cite{Stockton 2004,Mirrahimi 2007} that, for the ideal system subjected 
to the SME \eqref{SME} with $(H, M, L)=(J_y, \sqrt{\kappa}J_z, 0)$, the Dicke state 
$\ket{N/2, m}$ for arbitrary $m\in[-N/2, N/2]$ can be deterministically generated by 
an appropriate MBF control. 
Now using the lower bound $J_*$ we can evaluate how much this MBF control method could 
work under decoherence. 
Let $L=\sqrt{\gamma}J_-$. 
Then, the lower bound for $\ket{\psi}=\ket{l, m}$ is calculated as 
\begin{equation*}
     J_{\ast} = \frac{1}{2}
           \left[\frac{\gamma(l^{2}+l-m^{2}+m)}{2\kappa m^{2} 
               + 2\gamma(l^2+l-m^2)+\bar{u}\sqrt{l^2+l-m^2}} \right]^{2}.
\end{equation*}
Figure~3(a) shows the case of $N=20$ atoms, for the target Dicke state $\ket{\psi}=\ket{10, m}$. 
We observe that, as in the previous studies, the measurement drastically decreases $J_*$ 
especially for the state with large $|m|$, e.g., the CSS $\ket{10, 10}$. 
Meanwhile the lower bound at around the entangled state $\ket{10, 0}$ is almost unaffected 
by the measurement. 
Actually in general, for a Dicke state with large $|m|\lesssim l=N/2$, such as the CSS, the 
measurement term proportional to $\kappa$ is dominant in the denominator of $J_*$, while 
for highly entangled Dicke states with $m\sim 0$ the decoherence term proportional to $\gamma$ 
becomes dominant. 
In particular, 
\begin{align*}
     & J_*(\ket{N/2, N/2}) = 
                \left[ \frac{ \sqrt{2}\gamma N}
                               { \kappa N^2 + 2\gamma N + \bar{u}\sqrt{2N} } \right]^2, 
\\
     & J_*(\ket{N/2, 0}) = 
             \frac{1}{2} \left[
                 \frac{ \gamma(N^2+2N)}
                        { 2\gamma N^2 + 4\gamma N + 2\bar{u}\sqrt{N^2+2N}} \right]^2.
\end{align*}
Therefore, for a large ensemble limit $N\rightarrow \infty$, we have 
$J_*(\ket{N/2, N/2})\rightarrow 0$ and $J_*(\ket{N/2, 0})\rightarrow 1/8$. 
Note that this fundamental bound $J_*=1/8$ is applied to all highly entangled Dicke states 
satisfying $m\sim 0$ and $l \lesssim N/2 \gg 1$. 
That is, while no limitation appears for the case of the CSS thanks to the measurement effect, 
generating those highly entangled Dicke states is strictly prohibited, irrespective of the use of 
measurement and control. 
This result implies that, in practice, there exists a strict limitation in quantum magnetometry 
that utilizes a highly entangled Dicke state \cite{Duan NJP 2014}.

\begin{figure}[htbp]
\includegraphics[width=8.8cm]{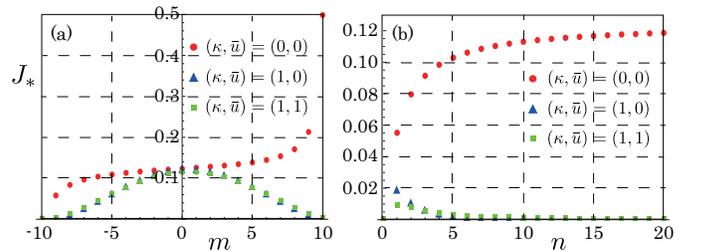}
\caption{
The lower bound $J_*$ for (a) the atomic Dicke states and (b) the optical Fock states, 
in the units of $\gamma=1$.}
\end{figure}

Another important subject in quantum metrology is the frequency standard, where the 
GHZ state $\ket{{\rm GHZ}}=(\ket{\uparrow}^{\otimes N}
+\ket{\downarrow}^{\otimes N})/\sqrt{2}$ 
is used for estimating the atomic frequency, over the standard quantum limit attained with 
the use of the product state 
$\ket{+}^{\otimes N}=(\ket{\uparrow}/\sqrt{2}+\ket{\downarrow}/\sqrt{2})^{\otimes N}$ 
\cite{Wineland 1996}. 
The main issue of this technique is that the estimation performance is severely limited 
\cite{Huelga 1997,Dorner 2012} due to the dephasing noise, which affects on both the state 
preparation process and the free-precession process. 
Here we characterize the performance degradation occurring in the former process, using the 
lower bound $J_*$. 
In the usual setup where no continuous monitoring is applied, the realistic system obeys the 
master equation $d\bar{\rho}/dt=-i[H, \bar{\rho}]+\mathcal{D}[L]\bar{\rho}$, where 
$L=\sqrt{\gamma}J_z$ represents the dephasing process and $H$ is a system Hamiltonian 
representing an open-loop control. 
Then the lower bounds for the above two states are given by 
\begin{align*}
     & J_*(\ket{+}^{\otimes N}) 
         =\left(\frac{\gamma N}{\sqrt{2}\gamma N 
               + \gamma\sqrt{6N^2-4N} + 4\,{\mathcal U}}\right)^2,
\\
     & J_{\ast}(\ket{{\rm GHZ}})
         =\left(\frac{\gamma N^2}{2\sqrt{2}\gamma N^2+4\,{\mathcal U}}\right)^2.
\end{align*}
Thus, irrespective of control, $J_*(\ket{+}^{\otimes N}) \rightarrow 1/(8+4\sqrt{3})$ and 
$J_*(\ket{{\rm GHZ}}) \rightarrow 1/8$ in the limit $N\to \infty$, under the assumption that 
${\mathcal U}$ is of the order at most $\sqrt{N}$ and $N$, respectively. 
Hence, the GHZ state is harder to prepare than the product one, and this gap would 
erase the quantum advantage obtained using the GHZ state in the ideal setting. 
In both cases, the estimation performance must be severely limited in the presence of 
decoherence, if the total time taken for state preparation and free-precession dynamics 
becomes long; 
thus, these two processes have to be carried out in as short a time as possible.


\subsection{Fock state}

The last case study is the problem of generating a Fock state in an optical cavity. 
In the setup of \cite{Geremia 2006,Yanagisawa 2006,Molmer 2007}, the conditional cavity 
state obeys the SME \eqref{SME} with $M=\sqrt{\kappa}a^\dagger a$ and 
$H=i(a^\dagger - a)$, where $a$ is the annihilation operator; 
then it was proven in the ideal case (i.e., $L=0$) that, by choosing an appropriate MBF 
input $u_t$, one can deterministically steer the state to a target Fock state $\ket{n}$. 
Now the lower bound $J_*$ can be used to evaluate the performance of this MBF control 
in the presence of decoherence. 
A typical decoherence is the photon leakage modeled by $L=\sqrt{\gamma}a$. 
In this setting, $J_*$ for $\ket{\psi}=\ket{n}$ is calculated as 
\[
     J_{\ast}(\ket{n})=\frac{1}{2}\left[
            \frac{\gamma n}{2\kappa n^2+\gamma (2n +1)+\bar{u}\sqrt{4n+2}} \right]^{2}.
\]
Figure~3(b) plots $J_{\ast}$ in the case $\gamma=1$. 
As in the previous studies, the measurement drastically decreases $J_*$. 
However, $J_*\sim 0$ for large Fock states $\ket{n}~(n \gtrsim 5)$ does not necessarily mean 
that those states can be exactly stabilized via MBF; 
rather a large Fock state might be hard to prepare compared to a small one such as $\ket{1}$. 
Hence, in this problem the lower bound only for small Fock states $\ket{n}~(n \lesssim 4)$ 
has a practical meaning.


\section{Conclusion}

In this paper, we have derived the general lower bound $J_*$ of the distance between 
the controlled quantum state under decoherence and an arbitrary target state. 
The lower bound can be straightforwardly calculated and used as a useful guide for 
engineering open quantum systems; 
for instance, in the reservoir engineering scenario, the system should be configured so that 
$J_*$ takes the minimum for a given target state. 
An important remaining work is to explore an achievable lower bound and develop an 
efficient method for synthesizing the controller (e.g., the MBF control input) that achieves 
the bound.

This work was supported in part by JST PRESTO Grant No. JPMJPR166A.


\appendix

\section{Proof of the theorem}

We prove the theorem in the MBF setting; the open-loop control and reservoir engineering 
case is obtained by simply setting the control signal $u_t$ to be independent of the 
conditional state $\rho_t$.

First, the infinitesimal change of the random variable 
\[
       j_t=1-{\rm Tr}(Q\rho_t), ~~~ Q=\ket{\psi}\bra{\psi}, 
\]
where $\rho_t$ is the solution of the stochastic master equation \eqref{SME} and 
$\ket{\psi}$ is the target, is given by 
\begin{eqnarray}
\label{Suppl SME}
& & \hspace*{-2em}
     dj_{t}= -{\rm Tr}(Qd\rho_{t})
\nonumber \\ & & \hspace*{-0.56em}
        =  -{\rm Tr}\big\{
                Q \big( -i[u_{t}H, \rho_{t}]dt + \mathcal{D}[L]\rho_{t}dt 
\nonumber \\ & & \hspace*{1em}
            \mbox{}  +\mathcal{D}[M]\rho_{t}dt + \mathcal{H}[M]\rho_{t}dW_{t} \big) \big\}
\nonumber \\ & & \hspace*{-0.57em}
      = {\rm Tr}\big( iu_{t}[Q, H]\rho_{t}\big)dt 
                 - {\rm Tr}\big(Q\mathcal{D}[L]\rho_{t}\big)dt  
\nonumber \\ & & \hspace*{1em}
         \hbox{}  - {\rm Tr}\big(Q\mathcal{D}[M]\rho_{t}\big)dt 
                 - {\rm Tr}\big(Q\mathcal{H}[M]\rho_{t}\big)dW_{t}. 
\end{eqnarray}
The classical stochastic process (the Wiener process) $W_t$ satisfies the Ito rule 
$dW_{t}^2=dt$ and $\mathbb{E}(W_t)=0$. The ensemble average of this equation, 
with respect to $W_t$, is thus 
\begin{eqnarray}
\label{Suppl ME}
& & \hspace*{-1em}
     \frac{d\mathbb{E}(j_t)}{dt} 
      = {\rm Tr}\big\{ i[Q, H]\mathbb{E}(u_t\rho_t)\big\}
                 -{\rm Tr}\big\{ Q\mathcal{D}[L]\mathbb{E}(\rho_{t}) \big\}
\nonumber \\ & & \hspace*{3em}
        \mbox{}  - {\rm Tr}\big\{ Q\mathcal{D}[M]\mathbb{E}(\rho_{t}) \big\}. 
\end{eqnarray}
Note again that $u_t$ is a function of $\rho_t$ in the context of MBF control.

In the proof we often use the Schwarz inequality for matrices (or bounded operators) 
$X$ and $Y$; 
\[
        \|X\|_{\rm F} \cdot \|Y\|_{\rm F} 
           \geq \frac{1}{2}\Big| {\rm Tr}(X^\dagger Y + Y^\dagger X) \Big|, 
\]
where $\| X \|_{\rm F}:=\sqrt{{\rm Tr}(X^\dagger X)}$ is the Frobenius norm. 
In particular, if $X$ and $Y$ are Hermitian, then 
$\|X\|_{\rm F} \cdot \|Y\|_{\rm F} \geq | {\rm Tr}(XY)|$ holds. 
The following inequality is also often used: 
\begin{eqnarray}
& & \hspace*{-1em}
      \|\rho_{t}-Q\|_{\rm F} = \sqrt{{\rm Tr}[(\rho_{t}-Q)^{2}]} 
              =\sqrt{ {\rm Tr} (\rho_{t}^{2}-2\rho_{t}Q+Q^{2})}
\nonumber \\ & & \hspace*{3.7em}
       \leq \sqrt{2-2{\rm Tr}(\rho_{t}Q)}=\sqrt{2j_{t}}, 
\nonumber
\end{eqnarray}
where ${\rm Tr}(\rho_t^2)\leq 1$ and ${\rm Tr}(Q^2)={\rm Tr}(Q)=1$ are used. 

We begin with calculating a lower bound of the first term on the rightmost side of 
Eq.~\eqref{Suppl SME} as follows: 
\begin{eqnarray*}
{\rm Tr}\big( iu_{t}[Q, H]\rho_{t} \big)  
     &\geq& -\bar{u} \Big| {\rm Tr}\big( i[Q, H]\rho_{t}\big) \Big| \\
     &=& -\bar{u} \Big| {\rm Tr}\big\{ i[Q, H] (\rho_t-Q)\big\} \Big| \\
     &\geq& -\bar{u} \|i[H, Q]\|_{\rm F} \cdot \|\rho_t-Q\|_{\rm F} \\
     &=&-\bar{u} \sqrt{ {\rm Tr}\{ (iHQ-iQH)^2 \} }\cdot \sqrt{2j_t} \\
     &\geq& -2\bar{u}\sqrt{ \bra{\psi}H^2\ket{\psi} - \bra{\psi}H\ket{\psi}^2 } \cdot \sqrt{j_{t}},
\end{eqnarray*} 
where $\bar{u}:=\max\{|u_{t}|\}$ is the upper bound of the control input. 
Then, by taking the ensemble average of this equation with respect to $W_t$, we have 
\begin{eqnarray}
& & \hspace*{-1em}
     {\rm Tr}\big\{ i[Q, H]\mathbb{E}(u_t\rho_t)\big\}
\nonumber \\ & & \hspace*{3em}
       \geq -2\bar{u}\sqrt{ \bra{\psi}H^2\ket{\psi} - \bra{\psi}H\ket{\psi}^2 } \cdot \mathbb{E}(\sqrt{j_{t}})
\nonumber \\ & & \hspace*{3em}
       \geq -2\bar{u}\sqrt{ \bra{\psi}H^2\ket{\psi} - \bra{\psi}H\ket{\psi}^2 } \cdot \sqrt{\mathbb{E}(j_t)},
\nonumber
\end{eqnarray}
where $\mathbb{E}(\sqrt{j_{t}}) \leq \sqrt{\mathbb{E}(j_{t})}$ is used. 
Next, the second term on the rightmost side of Eq.~\eqref{Suppl SME} can be lower 
bounded as follows; 
\begin{eqnarray}
& & \hspace*{-1em}
        -{\rm Tr}\big(Q\mathcal{D}[L]\rho_{t}\big)
              =-{\rm Tr}\Big[Q \big(
                       L\rho_{t}L^{\dagger} - \frac{1}{2}\rho_{t}L^{\dagger}L-\frac{1}{2}L^{\dagger}L\rho_{t}
                          \big)\Big]
\nonumber \\ & & \hspace*{-0.5em}
         = - {\rm Tr}\big[L^{\dagger}QL(\rho_{t}-Q)\big] - {\rm Tr}(L^{\dagger}QLQ)
            + {\rm Tr}(QL^{\dagger}LQ)
\nonumber \\ & & \hspace*{1em}
         \mbox{}
         -\frac{1}{2}{\rm Tr}\big[ (Q-\rho_{t})L^{\dagger}LQ 
              + (L^{\dagger}LQ)^\dagger (Q-\rho_{t}) \big]
\nonumber \\ & & \hspace*{-0.5em}
        \geq - \|L^{\dagger}QL \|_{\rm F} \cdot \|\rho_{t}-Q\|_{\rm F} 
              - \| L^\dagger L Q\|_{\rm F} \cdot \|\rho_{t}-Q\|_{\rm F}
\nonumber \\ & & \hspace*{1em}
        \mbox{}
         + {\rm Tr}(L^{\dagger}LQ)- {\rm Tr}(L^{\dagger}QLQ)
\nonumber \\ & & \hspace*{-0.5em}
        \geq - \Big\{ \sqrt{{\rm Tr}\big[(L^{\dagger}QL)^2\big]}
                 +\sqrt{{\rm Tr}\big[(L^{\dagger}LQ)^{\dagger}(L^{\dagger}LQ)\big]} \Big\}\sqrt{2j_{t}}
\nonumber \\ & & \hspace*{1em}
          \mbox{}
              + {\rm Tr}(L^{\dagger}LQ)-{\rm Tr}(L^{\dagger}QLQ)
\nonumber \\ & & \hspace*{-0.5em}
         = - \sqrt{2}\left( \bra{\psi} LL^\dagger \ket{\psi} 
                  + \sqrt{ \bra{\psi} (L^\dagger L)^2 \ket{\psi} }\right)
                           \sqrt{j_{t}}
\nonumber \\ & & \hspace*{1em}
          \mbox{}
               + \bra{\psi} L^\dagger L\ket{\psi} - |\bra{\psi}L\ket{\psi}|^2
\nonumber \\ & & \hspace*{-0.5em}
          = - \sqrt{2}\left( \| L^\dagger \ket{\psi} \|^2 + \|L^\dagger L\ket{\psi}\| \right)\sqrt{j_{t}} 
\nonumber \\ & & \hspace*{1em}
          \mbox{}
                +\| L\ket{\psi}\|^2  - |\bra{\psi}L\ket{\psi}|^2. 
\nonumber
\end{eqnarray}
Hence again it follows from $\mathbb{E}(\sqrt{j_{t}}) \leq \sqrt{\mathbb{E}(j_{t})}$ that 
\begin{eqnarray*}
& & \hspace*{-2em}
         - {\rm Tr}\big\{Q\mathcal{D}[L]\mathbb{E}(\rho_t)\big\}
\nonumber \\ & & \hspace*{0em}
         \geq 
             -\sqrt{2}\left( \| L^\dagger \ket{\psi} \|^2 
                 + \|L^\dagger L\ket{\psi}\| \right)\sqrt{\mathbb{E}(j_{t})} 
\nonumber \\ & & \hspace*{1em}
         \mbox{}                 
                 + \| L\ket{\psi}\|^2 - |\bra{\psi}L\ket{\psi}|^2. 
\end{eqnarray*} 
The same inequality as above, with $L$ replaced by $M$, holds. 
Hence, combining the above three inequalities with Eq.~\eqref{Suppl ME} and using the 
definition $J_{t}=\mathbb{E}(j_{t})=1- {\rm Tr}\big\{Q\mathbb{E}(\rho_t)\big\}
=1-{\rm Tr}(Q\bar{\rho}_{t})$, we end up with 
\begin{equation}
\label{Suppl EJ ineq}
      \frac{dJ_{t}}{dt} \geq 
           -\mathcal{U}\sqrt{J_{t}} - \mathcal{A}\sqrt{J_{t}} + \mathcal{E},
\end{equation}
where
\begin{eqnarray*}
     \frac{\mathcal{A}}{\sqrt{2}} &=& 
        \| L^\dagger \ket{\psi} \|^2 + \|L^\dagger L\ket{\psi}\| 
        +  \| M^\dagger \ket{\psi} \|^2 + \|M^\dagger M\ket{\psi}\| ,  
\\
     \mathcal{U} &=& 2\bar{u}\sqrt{ \bra{\psi}H^2\ket{\psi} - \bra{\psi}H\ket{\psi}^2 }, ~~
     \bar{u}:=\max\{|u_{t}|\}, 
\\
     \mathcal{E} &=& 
        \| L\ket{\psi}\|^2 - |\bra{\psi}L\ket{\psi}|^2 + \| M\ket{\psi}\|^2 - |\bra{\psi}M\ket{\psi}|^2.
\end{eqnarray*}

To obtain the lower bound of $J_t$ in the limit $t\rightarrow\infty$, let us consider 
the function $f(x)=-\mathcal{U}\sqrt{x}-\mathcal{A}\sqrt{x}+\mathcal{E}$ in the range $x\in[0,1]$. 
Clearly, $f(x)$ is a monotonically decreasing function with respect to $x$. 
Also, from the Schwarz inequality $\| L\ket{\psi}\|^2 - |\bra{\psi}L\ket{\psi}|^2 \geq 0$, 
we have $f(0)=\mathcal{E}\geq0$. 
Moreover, $f(1)=\mathcal{E}-\mathcal{A}-\mathcal{U}\leq0$ holds, because 
\begin{eqnarray*}
 \| L\ket{\psi}\|^2&=&{\rm Tr}(L^{\dag}LQ) \\
&=&\frac{1}{2}\Big\{ {\rm Tr}\big[(L^{\dag}LQ)^\dagger Q\big] 
+ {\rm Tr}\big[Q^\dagger (L^{\dag}LQ)\big] \Big\}  \\
       &\leq& \| L^{\dag}LQ \|_{\rm F} \cdot \| Q \|_{\rm F} 
       = \| L^{\dag}LQ \|_{\rm F}
       = \| L^\dagger L \ket{\psi} \|,
\end{eqnarray*}
which clearly leads to $\mathcal{E}-\mathcal{A}\leq 0$ and accordingly 
$\mathcal{E}-\mathcal{A}-\mathcal{U}\leq0$. 
In what follows we consider the case ${\mathcal E}>0$. 
Then, from the above properties of $f(x)$, the equation $f(J_*)=0$ has a unique solution 
$J_*$ in $(0,1]$. 
Now suppose that $J_{\tau}<J_*$ at a given time $\tau$; 
then Eq.~\eqref{Suppl EJ ineq} leads to 
\begin{eqnarray*}
     \frac{dJ_{t}}{dt}\Big|_{t=\tau} &\geq& 
           -\mathcal{U}\sqrt{J_{\tau}} - \mathcal{A}\sqrt{J_{\tau}} + \mathcal{E} \\
           &>& -\mathcal{U}\sqrt{J_*} - \mathcal{A}\sqrt{J_*} + \mathcal{E}
           = 0. 
\end{eqnarray*}
This means that $J_{t}$ locally increases in time for $t\geq \tau$. 
Because this argument is true for any $\tau$ such that the inequality $J_{\tau}<J_*$ 
holds, $J_{t}$ increases until $J_{t}$ coincides with $J_*$; 
i.e., $\lim_{t\rightarrow\infty} J_{t}=J_*$. 
On the other hand for the range such that $J_{\tau}\geq J_*$ the inequality 
\eqref{Suppl EJ ineq} does not say anything about the local time evolution of $J_{t}$ 
for $t\geq \tau$. 
As a result, in the long time limit we have 
\[
     \lim_{t\rightarrow\infty}J_{t} 
       \geq J_{\ast}=\left(\frac{\mathcal{E}}{\mathcal{A}+\mathcal{U}} \right)^2. 
\]
Note that this inequality is valid for the case $\mathcal{E}=0$ as well.

Next we prove that $J_t \geq J_*$ holds for all $t\in[t_{0},\infty)$, if the initial value 
$J_{t_0}$ is bigger than $J_*$. 
For the proof we use the following fact; see e.g., \cite{Diff}. 
\\

{\it Theorem~2}: 
Consider the following real-valued 1-dimensional ordinary differential equation:
\begin{eqnarray*}
     \frac{dx(t)}{dt}=f(x(t)), \ \ t\in[t_{0},\infty).
\end{eqnarray*}
If $dx_1(t)/dt \leq f(x_1(t))$ and $f(x_2(t)) \leq dx_2(t)/dt$,~$\forall t\in[t_{0},\infty)$ hold 
for the initial values satisfying $x_1(t_0) \leq x_2(t_0)$, then $x_{1}(t) \leq x_{2}(t)$ holds for 
$\forall t\in[t_{0},\infty)$.
\\

{\it Proof:}
A contradiction argument will be used.
Suppose that there exists $t\in[t_{0},\infty)$ such that  $x_{1}(t) > x_{2}(t)$. 
Then, because $x_1(t_0) \leq x_2(t_0)$, there exists $T\geq t_{0}$ satisfying $x_{1}(T)=x_{2}(T)$. 
Moreover, there exists $h>0$ such that $x_{1}(T+h) > x_{2}(T+h)$ holds. 
Hence, 
\begin{eqnarray*}
     \frac{dx_1(t)}{dt}\Big|_{T+0} 
         &=& \lim_{h\to +0}\frac{x_1(T+h)-x_1(T)}{h} \\
         &>& \lim_{h\to +0}\frac{x_2(T+h)-x_2(T)}{h}=\frac{dx_2(t)}{dt} \Big|_{T+0}.
\end{eqnarray*}
Then, from the assumption of Theorem~2, 
\begin{eqnarray*}
      f\left(x_1(T)\right) 
        \geq \frac{dx_{1}(t)}{dt}\Big|_{T+0} 
        >       \frac{dx_{2}(t)}{dt} \Big|_{T+0}
        \geq f\left(x_{2}(T)\right).
\end{eqnarray*}
This is a contradiction to $x_{1}(T)=x_{2}(T)$. 
Therefore $x_{1}(t) \leq x_{2}(t)$ holds for $\forall t\in[t_{0},\infty)$. 
$\blacksquare$
\\

Let us apply Theorem 2 to the case 
$f(x)=-\mathcal{U}\sqrt{x}-\mathcal{A}\sqrt{x}+\mathcal{E}$. 
Assuming that $x_1(t_0)=J_*=\left\{\mathcal{E}/(\mathcal{A}+\mathcal{U})\right\}^{2}$,  
we have $dx_1(t)/dt=f(x_1(t))=0$ and $x_1(t)=x_1(t_0)=J_*$ for $\forall t\in[t_{0},\infty)$. 
Also we take $x_2(t)=J_{t}$, which satisfies the inequality \eqref{Suppl EJ ineq}, 
i.e., $dx_2(t)/dt \geq f(x_2(t))$. 
Thus, from Theorem~2, if $J_* = x_1(t_0) \leq x_2(t_0) = J_{t_0}$, 
then $J_*=x_1(t) \leq x_2(t) = J_{t}$ for all $t\in[t_{0},\infty)$.
That is, we obtain 
\[
    J_t \geq J_{\ast}=\left(\frac{\mathcal{E}}{\mathcal{A}+\mathcal{U}} \right)^2, ~~~
       \forall t\in[t_{0}, \infty). 
\]
This is end of the proof of Theorem~1.


\section{Generalization of the theorem}

If the system dynamics is validly modeled by the stochastic master equation
\begin{eqnarray*}
          d\rho_{t} 
             = &-&i \Big[ \sum_{j}u_{j,t}H_{j}, \rho_{t} \Big]dt+\sum_{j}\mathcal{D}[L_{j}]\rho_{t}dt \\
                           &+&\sum_{j}\mathcal{D}[M_{j}]\rho_{t}dt+\sum_{j}\mathcal{H}[M_{j}]\rho_{t}dW_{t}, 
\end{eqnarray*}
for the MBF case or the master equation 
\begin{equation*}
      \frac{d\bar{\rho}_t}{dt}
         = -i \Big[ \sum_{j}u_{j,t} H_{j}, \bar{\rho}_t \Big] 
           + \sum_{j}\mathcal{D}[L_{j}] \bar{\rho}_t + \sum_{j}\mathcal{D}[M_{j}] \bar{\rho}_t, 
\end{equation*}
for the open-loop control or reservoir engineering case, by the straightforward extension 
of the above discussion we find that the lower bound is given by 
$J_*={\mathcal E}^2/({\mathcal A}+{\mathcal U})^2$ with 
\begin{eqnarray*}
     \mathcal{A}&=& \sqrt{2} \sum_j
           \left( \|L_j^{\dag}\ket{\psi}\|^{2} + \|L_j^{\dag}L_j\ket{\psi}\| \right) \\
& &+\sqrt{2} \sum_j
           \left( \|M_j^{\dag}\ket{\psi}\|^{2} + \|M_j^{\dag}M_j\ket{\psi}\| \right), 
\\
     \mathcal{U}&=&2\sum_j\ \bar{u}_{j} 
           \sqrt{\bra{\psi}H_{j}^2\ket{\psi}-\bra{\psi}H_{j}\ket{\psi}^{2}}, 
\\
       && ~~~~ \bar{u}_j = {\rm max}\{ |u_{j,t}| \},
\\
      \mathcal{E}&=&
            \sum_j \left(\left\|L_j\ket{\psi}\right\|^{2}-\left|\bra{\psi}L_j\ket{\psi}\right|^{2} \right) \\
    & &+\sum_j \left(\left\|M_j\ket{\psi}\right\|^{2}-\left|\bra{\psi}M_j\ket{\psi}\right|^{2} \right). 
\end{eqnarray*}
Using this result we can involve a fixed system Hamiltonian in addition to controllable 
Hamiltonians; 
in the simple case where the Hamiltonian is given by $H_0 + u_t H_1$ with $H_0$ a fixed 
system Hamiltonian, we have 
\begin{eqnarray*}
     {\mathcal U} 
        &=& 2 \sqrt{ \bra{\psi}H_0^2\ket{\psi} - \bra{\psi}H_0\ket{\psi}^2 }
\\
        &+& 2 \bar{u} \sqrt{ \bra{\psi}H_1^2\ket{\psi} - \bra{\psi}H_1\ket{\psi}^2 }, 
\end{eqnarray*}
with $\bar{u}=\max\{ |u_t| \}$.


\section{A measurement-based feedback control law}

Let us consider the continuously-monitored system whose dynamics is given by 
the following SME:
\begin{equation*}
      d\rho_{t}=-i[u_{t}J_{y}, \rho_{t}]dt+\mathcal{D}[J_{z}]\rho_{t}dt+\mathcal{H}[J_{z}]\rho_{t}dW_{t},
\end{equation*}
where $J_y$ and $J_z$ are the angular momentum operators. 
Now the target state $\ket{\psi}$ is set to be one of the eigenstates of $J_{z}$. 
In Ref.~\cite{Mirrahimi 2007} the authors proposed the following feedback control law that 
deterministically steers the conditional state $\rho_t$ to the target state: 
\par
\begin{enumerate}
\item $u_{t}=-\alpha {\rm Tr}\left\{ i[J_{y}, \rho_{t}]Q \right\}$ if ${\rm Tr}(Q\rho_{t})\geq\beta$, 
\item $u_{t}=\alpha$ if ${\rm Tr}(Q \rho_{t})\leq\beta/2$, 
\item If $\rho_{t} \in \mathcal{B}=\{ \rho~|~\beta/2<{\rm Tr}(Q\rho_{t})<\beta \}$, then 
$u_{t}=-\alpha{\rm Tr}\left\{ i[J_{y}, \rho_{t}]Q\right\}$ in the case $\rho_{t}$ last entered 
$\mathcal{B}$ through the boundary ${\rm Tr}(Q\rho)=\beta$, and $u_{t}=\alpha$ otherwise, 
\end{enumerate}
where $\alpha$ and $\beta$ are positive constants. 
In fact, it was proven that there exists $\beta>0$ such that 
$\rho_{t} \rightarrow Q=\ket{\psi}\bra{\psi}$ almost surely.

Here we calculate the upper bound $\bar{u}$ of the above MBF control input, in the qubit 
control problem discussed in Sec. III~A. 
First we have 
\begin{eqnarray*}
        {\rm Tr}(\Delta \sigma_{y}^{2}\rho_{t}) 
          &=& {\rm Tr}(\sigma_{y}^{2}\rho_{t})-[{\rm Tr}(\sigma_{y}\rho_{t})]^{2} \\
          &=& 1-[{\rm Tr}(\sigma_{y}\rho_{t})]^{2} \leq 1, 
\\
        {\rm Tr}(\Delta Q^{2}\rho_{t}) 
          &=& {\rm Tr}(Q^{2}\rho_{t})-[{\rm Tr}(Q\rho_{t})]^{2} \\
           &=&{\rm Tr}(Q\rho_{t})-[{\rm Tr}(Q\rho_{t})]^{2} \\
           &=&(1-J_{t})J_{t}\leq\frac{1}{4}.
\end{eqnarray*}
Then, the Robertson inequality 
${\rm Tr}(\Delta \sigma_{y}^{2}\rho_{t}){\rm Tr}(\Delta Q^{2}\rho_{t}) \geq 
|{\rm Tr}\{ i[\sigma_{y}, Q]\rho_{t}\}|^{2}/4$ leads to 
\begin{eqnarray*}
     \Big|-\alpha {\rm Tr}\left\{ i[J_{y}, \rho_{t}]Q \right\} \Big| 
       &\leq& 2\alpha \sqrt{{\rm Tr}(\Delta \sigma_{y}^{2}\rho_{t})}
                           \sqrt{{\rm Tr}(\Delta Q^{2}\rho_{t})}  \\
       &\leq& 2\alpha \cdot\frac{1}{2}=\alpha.
\end{eqnarray*}
Hence, together with the other case of input, we have $\bar{u}=\max\{|u_t|\}=\alpha$; 
in the numerical simulation depicted in Fig.~1(c) in the main text, $\bar{u}=\alpha=1$ 
was chosen.


\section{Detailed calculations of the lower bound}


\subsection{Qubit}

The target state is $\ket{\psi}=[{\rm cos}\theta,\ e^{i\varphi}{\rm sin}\theta]^{\top}$ 
$(0\leq \theta< \pi/2, 0\leq \varphi< 2\pi)$, and the system operators are 
$H=\sigma_{y}$, $M=\sqrt{\kappa}\sigma_{z}$, and $L=\sqrt{\gamma}\sigma_{-}$. 
Then we have 
\begin{eqnarray*}
    \frac{\mathcal{A}}{\sqrt{2}} &=& 
         \|L^{\dagger}\ket{\psi}\|^{2} + \|L^{\dagger}L\ket{\psi}\|
                        + \|M^{\dagger}\ket{\psi}\|^{2} + \|M^{\dagger}M\ket{\psi}\| \\
     &=& 2\kappa +\gamma \sin^2\theta + \gamma \cos\theta , \\
    \mathcal{U} 
     &=& 2\bar{u} \sqrt{ \bra{\psi} H^2 \ket{\psi} - \bra{\psi}H\ket{\psi}^{2}} \\
     &=& 2\bar{u}\sqrt{ 1-\sin^{2}2\theta \sin^{2}\varphi }, 
\\
    \mathcal{E} &=& \| L\ket{\psi}\|^{2} - |\bra{\psi}L\ket{\psi}|^{2} 
             + \|M\ket{\psi}\|^{2} - |\bra{\psi}M\ket{\psi}|^{2} \\
            &=& \kappa \sin^{2} 2\theta + \gamma \cos^{4}\theta.
\end{eqnarray*}
%


\subsection{Qutrit}

The target state is limited to the real vector 
$\ket{\psi}=[ \sin(\theta/2)\cos(\varphi/2), \, \cos(\theta/2), \, \sin(\theta/2) \sin(\varphi/2) ]^\top$. 
The system operators are given by 
\begin{eqnarray*}
H&=&\frac{1}{\sqrt{2}}\left[\begin{array}{rrr}
0 & -i & 0  \\
i & 0 & -i \\
0 & i &  0
\end{array}\right], \ 
M=\sqrt{\kappa}\left[\begin{array}{rrr}
1 & 0 & 0  \\
0 & 0 & 0 \\
0 & 0 & -1  
\end{array}\right], \\ 
L&=&\sqrt{\gamma}\left[\begin{array}{rrr}
0 & 0 & 0  \\
1 & 0 & 0 \\
0 & 1 & 0  
\end{array}\right].
\end{eqnarray*}
Then, from the definition of ${\mathcal A}$, ${\mathcal U}$, and ${\mathcal E}$, we have 
\begin{eqnarray*}
& & \hspace*{-1em}
  \frac{{\mathcal A}}{\sqrt{2}}
      =\kappa\left({\rm sin}^{2}\frac{\theta}{2}+{\rm sin}\frac{\theta}{2}\right)
\nonumber \\ & & \hspace*{-0.5em}
      \mbox{}
      + \gamma\left({\rm cos}^{2}\frac{\theta}{2}
      +{\rm sin}^{2}\frac{\theta}{2}{\rm sin}^{2}\frac{\varphi}{2}
      + \sqrt{{\rm sin}^{2}\frac{\theta}{2}{\rm cos}^{2}\frac{\varphi}{2}
      +{\rm cos}^{2}\frac{\theta}{2}}\right), 
\nonumber \\ & & \hspace*{-1em}
    \mathcal{U}
      =\sqrt{2}\bar{u}\sqrt{1+{\rm cos}^{2}\frac{\theta}{2}-{\rm sin}^{2}\frac{\theta}{2}{\rm sin}\varphi}, 
\nonumber \\ & & \hspace*{-1em}
   \mathcal{E}
      =\kappa\left({\rm sin}^{2}\frac{\theta}{2}+{\rm sin}\frac{\theta}{2}\right)
\nonumber \\ & & \hspace*{-0.5em}
    \mbox{}  
      +\gamma\left({\rm sin}^{2}\frac{\theta}{2}{\rm sin}^{2}\frac{\varphi}{2}
      +{\rm cos}^{2}\frac{\theta}{2}
      +\sqrt{{\rm sin}^{2}\frac{\theta}{2}{\rm cos}^{2}\frac{\varphi}{2}+{\rm cos}\frac{\theta}{2}}\right).
\end{eqnarray*} 
%


\subsection{Dicke state}

The target is the Dicke state $\ket{l, m}$, and the system operators are $H=J_{y}$, 
$M=\sqrt{\kappa}J_{z}$, and $L=\sqrt{\gamma}J_-$. 
Recall that $\ket{l, m}$ is a common eigenvector of $J_z$ and the orbital angular momentum 
operator ${\bf J}^{2}=J_{x}^{2}+J_{y}^{2}+J_{z}^{2}$; 
\[
       J_{z}\ket{l, m}=m\ket{l, m},  ~~ {\bf J}^{2}\ket{l, m}=l(l+1)\ket{l, m}.
\]
Also the raising and lowering operators $J_{\pm}=J_{x}\pm iJ_{y}$ act on $\ket{l, m}$ 
as follows:
\[
     J_{\pm}\ket{l,m}=\sqrt{(l\mp m)(l\pm m+1)}\ket{l, m\pm1}.
\]
The following equations are also used; 
\begin{eqnarray*}
      J_{+}J_{-}&=&(J_{x}+iJ_{y})(J_{x}-iJ_{y})\\
                  &=&J_{x}^{2}+J_{y}^{2}-iJ_{x}J_{y}+iJ_{y}J_{x} \\
          &=&{\bf J}^{2}-J_{z}^{2}-i[J_{x}, J_{y}]={\bf J}^{2}-J_{z}^{2}+J_{z}, 
\end{eqnarray*}
and likewise $J_{-}J_{+} = {\bf J}^{2}-J_{z}^{2}-J_{z}$. 
Then we have 
\begin{eqnarray*}
& & \hspace*{-1em}
     \mathcal{A} 
         = \sqrt{2}\Big(\kappa\|J_{z}\ket{l,m}\|^{2}+\kappa\|J^2_{z}\ket{l,m}\|
\nonumber \\ & & \hspace*{2em}
      \mbox{}
         +\gamma\|J_{-}^{\dag}\ket{l,m}\|^{2}+\gamma\|J_{-}^{\dag}J_{-}\ket{l,m}\| \Big)
\nonumber \\ & & \hspace*{0em}
       = \sqrt{2}\Big[\kappa m^{2}+\kappa m^{2}
              +\gamma\{ l(l+1)-m^2-m\}       
\nonumber \\ & & \hspace*{2em}
      \mbox{}
          +\gamma\{l(l+1)-m^2+m\}\Big]  
\nonumber \\ & & \hspace*{0em}
       = \sqrt{2}\left\{2\kappa m^{2}+2\gamma(l^2+l-m^2)\right\}, 
\nonumber \\ & & \hspace*{-1em}
     \mathcal{U} 
        = \bar{u}\sqrt{ \left\|(J_{-}^{\dag}-J_{-})\ket{l,m}\right\|^{2}
              - \left|\bra{l,m}(J_{-}^{\dag}-J_{-})\ket{l,m}\right|^{2}}
\nonumber \\ & & \hspace*{0em}
     = \bar{u}\sqrt{ \left\{(l-m)(l+m+1)+(l+m)(l-m+1)\right\} }
\nonumber \\ & & \hspace*{0em}
     = \sqrt{2}\bar{u}\sqrt{l^2+l-m^2},
\nonumber \\ & & \hspace*{-0.9em}
      \mathcal{E} 
          = \kappa\|J_{z}\ket{l,m}\|^{2}-\kappa\left|\bra{l,m}J_{z}\ket{l,m}\right|^{2} 
\nonumber \\ & & \hspace*{2em}
      \mbox{}          
          + \gamma\|J_{-}\ket{l,m}\|^{2}-\gamma\left|\bra{l,m}J_{-}\ket{l,m}\right|^{2} 
\nonumber \\ & & \hspace*{0em}
        =\kappa m^{2}- \kappa m^{2}+\gamma (l+m)(l-m+1)
\nonumber \\ & & \hspace*{0em}        
        =\gamma(l^{2}+l-m^{2}+m).
\end{eqnarray*} 
%


\subsection{Product state of the spin superposition}

The target is the product state of the spin superposition, $\ket{+}^{\otimes N}$, 
where $\ket{+}=(\ket{\uparrow}+\ket{\downarrow})/\sqrt{2}$. 
The system is driven by the dephasing process $L=\sqrt{\gamma}J_{z}$ and an appropriate 
control Hamiltonian $H$, but it is not subjected to continuous monitoring and subsequent 
MBF (i.e., $M=0$). 
Recall that $J_z$ can be represented as 
\begin{equation}
\label{Jz representation}
     J_z = \frac{1}{2}\sum^{N}_{j=1}\sigma^{(j)}_z
            = \frac{1}{2}\sum^{N}_{j=1}(I\otimes \cdots \otimes \sigma_z \otimes \cdots \otimes I),
\end{equation}
where the Pauli matrix $\sigma_z$ appears in the $j$th component. 
$\sigma_z$ acts on $\ket{\uparrow}$ and $\ket{\downarrow}$ as 
$\sigma_z\ket{\uparrow}=\ket{\uparrow}$ and $\sigma_z\ket{\downarrow}=-\ket{\downarrow}$, 
respectively. 
Hence we have $\sigma_z\ket{+}=\ket{-}$, where 
$\ket{-}=(\ket{\uparrow}-\ket{\downarrow})/\sqrt{2}$. 
Moreover, 
\[
      L\ket{+}^{\otimes N} 
        = \frac{\sqrt{\gamma}}{2}\sum^{N}_{j=1} \ket{+}\cdots\ket{+}\ket{-}\ket{+}\cdots\ket{+}
\]
holds, where $\ket{-}$ appears in the $j$th component. 
This leads to $\bra{+}^{\otimes N} L\ket{+}^{\otimes N} =0$ and thus 
\[
      \mathcal{E} = \left\| L \ket{+}^{\otimes N} \right\|^2 
               - \left| \bra{+}^{\otimes N} L\ket{+}^{\otimes N} \right|^2 = \frac{\gamma N}{4}.
\]
Also, we have 
\begin{eqnarray*}
& & \hspace*{-1em}
    L^\dagger L \ket{+}^{\otimes N}
\nonumber \\ & & \hspace*{-0.5em}
        =\frac{\gamma}{4}\Big( N\ket{+}^{\otimes N} 
            +2\sum_{i\neq j} \ket{+}\ldots \ket{-} \cdots \ket{-} \ldots \ket{+} \Big),
\end{eqnarray*} 
where $\ket{-}$ appears only in the $i$th and $j$th components ($i\neq j$). 
Hence, 
\begin{eqnarray*}
        \mathcal{A} 
             &=& \sqrt{2}\left( \left\| L^{\dag} \ket{+}^{\otimes N} \right\|^{2} 
                    + \left\|L^{\dag}L \ket{+}^{\otimes N} \right\|\right) \\
        &=& \frac{\gamma}{4} \Big( \sqrt{2} N + \sqrt{6N^2-4N} \Big).  
\end{eqnarray*}
%


\subsection{GHZ state}

The target is 
$\ket{{\rm GHZ}}=\left(\ket{\uparrow}^{\otimes N}+\ket{\downarrow}^{\otimes N}\right)/\sqrt{2}$. 
As in the previous case, the system is driven by the dephasing process $L=\sqrt{\gamma}J_{z}$ 
and an appropriate Hamiltonian $H$, while not subjected to MBF (i.e., $M=0$). 
Using the representation \eqref{Jz representation} we have 
\begin{eqnarray*}
& & \hspace*{-1em}
     \frac{\mathcal{A}}{\sqrt{2}} 
         = \left\| L^{\dag}\ket{{\rm GHZ}}\right\|^{2}+\left\|L^{\dag}L\ket{{\rm GHZ}}\right\| 
\nonumber \\ & & \hspace*{-0.5em}
     =\frac{\gamma}{8}N^{2}\left\|\ket{\uparrow}^{\otimes N}-\ket{\downarrow}^{\otimes N}\right\|^{2}
     +\frac{\gamma}{4\sqrt{2}}N^{2}\left\|\ket{\uparrow}^{\otimes N}+\ket{\downarrow}^{\otimes N}\right\|
\nonumber \\ & & \hspace*{-0.5em}
     =\frac{\gamma}{4}N^{2}+\frac{\gamma}{4}N^{2} = \frac{\gamma}{2}N^{2}, 
\nonumber \\ & & \hspace*{-1em}
     \mathcal{E} = \left\|L\ket{{\rm GHZ}}\right\|^2 - \left|\bra{{\rm GHZ}}L\ket{{\rm GHZ}}\right|^2 
\nonumber \\ & & \hspace*{-0.2em}
     =\frac{\gamma}{8} N^2 \left\|\ket{\uparrow}^{\otimes N}-\ket{\downarrow}^{\otimes N}\right\|^2-0
     =\frac{\gamma}{4}N^2.
\end{eqnarray*} 
%


\subsection{Fock State}

The target is an arbitrary Fock state $\ket{n}$ and the system operators are 
given by $H=i(a^{\dag}-a)$, $M=\sqrt{\kappa}a^{\dag}a$, and $L=\sqrt{\gamma}a$. 
Using $a\ket{n}=\sqrt{n}\ket{n-1}$ and $a^{\dag}\ket{n}=\sqrt{n+1}\ket{n+1}$, we have 
\begin{eqnarray*}
& & \hspace*{-1em}
    \mathcal{A} 
      = \sqrt{2}\Big( \kappa\left\|a^{\dag}a\ket{n}\right\|^{2}
      +\kappa\left\|a^{\dag}a a^{\dag}a\ket{n}\right\|
\nonumber \\ & & \hspace*{3em}
    \mbox{}
        + \gamma\left\|a^{\dag}\ket{n}\right\|^{2}+\gamma\left\|a^{\dag}a\ket{n}\right\| \Big)
\nonumber \\ & & \hspace*{0em}
      = 2\sqrt{2}\kappa n^{2} + \sqrt{2}\gamma(2n+1),  
\nonumber \\ & & \hspace*{-1em}
    \mathcal{U}
      = 2\bar{u}\sqrt{\left\|i(a^{\dag}-a)\ket{n}\right\|^{2}-\bra{n}i(a^{\dag}-a)\ket{n}^{2}} 
\nonumber \\ & & \hspace*{0em}
      = 2\bar{u}\sqrt{2n+1},   
\nonumber \\ & & \hspace*{-1em}
    \mathcal{E} 
      = \kappa\left\|a^{\dag}a\ket{n}\right\|^{2}-\kappa\left|\bra{n}a^{\dag}a\ket{n}\right|^{2}
\nonumber \\ & & \hspace*{3em}
    \mbox{}
          +\gamma\left\|a\ket{n}\right\|^{2}-\gamma\left|\bra{n}a\ket{n}\right|^{2}  
\nonumber \\ & & \hspace*{-0.2em}
      = \gamma n. 
\end{eqnarray*} 
%


\end{document}